\documentclass[pdftex,twocolumn,epjc3]{svjour3}          

\smartqed  
\RequirePackage{graphicx}

\usepackage[numbers,sort&compress]{natbib}

\usepackage[utf8]{inputenc}
\usepackage[english]{babel}
\usepackage{amsmath,amsbsy,amsfonts,amsmath,bbm,latexsym,amssymb,bm}
\usepackage{caption}
\usepackage{subcaption}
\usepackage{feynmp-auto}
\usepackage[plainpages=false,pdfborder={0 0 0},linkbordercolor={1. 1. 1.},citebordercolor={1. 1. 1.},urlbordercolor={1. 1. 1.}]{hyperref}
\usepackage{cuted}

\newcommand{\be}{\begin{equation}}
\newcommand{\ee}{\end{equation}}
\newcommand{\ba}{\begin{Eqnarray}}
	\newcommand{\ea}{\end{Eqnarray}}
\def\bea{\begin{eqnarray}}
\def\eea{\end{eqnarray}}
\newcommand{\no}{\nonumber\\}

\newcommand{\bs}{\begin{subequations}}
	\newcommand{\es}{\end{subequations}}

\newcommand{\ie}{\textit{i.e.}\ }

\newcommand{\viz}{\textit{viz.}\ }

\renewcommand{\eqref}[1]{eq.~(\ref{#1})}

\def\n{\noindent}

\def\hs{\hspace}
\usepackage{color}
\usepackage{soul,xcolor}
\definecolor{orange}{rgb}{1,0.5,0}

\setstcolor{red}

\def\makeheadbox{
\vspace{-3mm}
{\raggedleft
{CFTP/20-005}\\
{KA-TP-03-2021}\\
{P3H-21-015}\\}}
\def\rubric{{}}

\begin{document}

\title{Leaks of CP violation in the real two-Higgs-doublet model}

\author{Duarte Fontes\thanksref{e1,addr1} \and
			Maximilian Löschner\thanksref{e2,addr2} \and 
			Jorge C. Rom\~{a}o\thanksref{e3,addr1} \and
			Jo\~{a}o P. Silva\thanksref{e4,addr1}  } 

\thankstext{e1}{e-mail: duartefontes@tecnico.ulisboa.pt}
\thankstext{e2}{e-mail: maximilian.loeschner@kit.edu}
\thankstext{e3}{e-mail: jorge.romao@tecnico.ulisboa.pt}
\thankstext{e4}{e-mail: jpsilva@cftp.ist.utl.pt}
\institute{Departamento de F\'{\i}sica and Centro de F\'{\i}sica Te\'{o}rica de
				Part\'{\i}culas (CFTP),\\
				\quad Instituto Superior T\'{e}cnico (IST), U. de Lisboa (UL),\\ 
				\quad Av. Rovisco Pais, P-1049-001 Lisboa, Portugal.\label{addr1} \and
				Institute for Theoretical Physics, Karlsruhe Institute for Technology,\\
				Wolfgang-Gaede-Straße 1, 76131 Karlsruhe, Germany \label{addr2}
}

\date{\today}

\maketitle

\begin{abstract}
We discuss the $Z_2$ symmetric two-Higgs-dou- blet model with a real
soft-breaking term (real 2HDM).
We explain in detail why it is not tenable to assume CP conservation
in the scalar sector to keep the dimension two term real,
while CP is violated by the dimension four Yukawa couplings.
We propose the calculation of the infinite
tadpole of the (would-be) pseudoscalar neutral scalar.
We construct a simple toy model with the same flaws,
where the unrenormalizable infinity is easier to calculate.
We then consider the same tadpole in the real 2HDM.
We spearhead this effort focusing on diagrams involving solely
bare quantities.
This involves hundreds of Feynman three-loop
diagrams that could feed the CP violation from the quark into the
scalar sector, and is only possible with state of the art
automatic computation tools.
Remarkably, some intermediate results agree when using three independent
derivations, including the peculiar cancellation of the leading pole
divergence due to a subtle interplay between masses and the Jarlskog
invariant, which we calculate analytically.
The calculation is not complete however,
since the full two-loop renormalization of the real 2HDM is not
yet available in the literature.
Still, we argue convincingly that there is an irremovable infinity.
\end{abstract}


\section{Introduction}
\label{sec:intro}

The exact implementation of the symmetry breaking sector of the gauge theory
describing fundamental particles is one of the most interesting open problems.
In particular, the recent discovery at LHC of a $125\, \textrm{GeV}$
fundamental scalar ($h_{125}$) \cite{Aad:2012tfa,Chatrchyan:2012xdj}
begs the question of how many
fundamental scalars there are in Nature.
The Standard Model (SM) contains one single scalar doublet,
but there is no fundamental requirement for this choice.

Indeed,
there have been countless studies on models with two or more Higgs doublets;
for reviews see \cite{hhg,Branco:2011iw,Ivanov:2017dad} and references therein.
The most studied model includes two Higgs doublets (2HDM) with a
$Z_2$ symmetry, softly broken by a real parameter,
with the additional requirement that the vacuum expectation values (vev)
are taken as real.
As such, the scalar sector by itself is CP conserving.
The $Z_2$ symmetry is carried over to the fermions in such a way
that only one scalar couples to all fermions of a given charge.
The crucial point is that the experimentally observed CP
violation~\cite{Zyla:2020zbs}
is accounted for by complex Yukawa couplings.
This is sometimes referred to as the ``real 2HDM'', because the
soft $Z_2$ breaking parameter is taken as real.
Remarkably,
this most prevalent model
	can suffer from from theoretical inconsistencies regarding renormalization,
a fact that is mostly ignored.\footnote{We know of only one
published exception, appearing in one line on page
4 of \cite{Altenkamp:2017ldc}.}
This is the problem we address in detail here.

The paper is organized as follows.
In section~\ref{sec:short}, we discuss the inconsistency of requiring CP conservation in the potential of the real 2HDM, while allowing for CP violation elsewhere.
We argue that, at sufficient high order in perturbation theory, there
	could
be divergences in CP-violating one-point and two-point functions that
	one can not remove
by the counterterms provided by the theory.
We show in section~\ref{sec:toymodel} that this is precisely what happens in a toy model suffering from the same inconsistency as the real 2HDM.
Then, in section 4, we present the result for the leading pole of the three-loop one-point function of the alleged CP-odd physical field of the real 2HDM.
We describe in section~\ref{sec:details} the details of the different steps involved in the three-loop calculation.
In section~\ref{sec:conclusions}, we summarize our conclusions.

\section{Shortcomings of the real 2HDM}
\label{sec:short}

Let us consider a $SU(2)_L \otimes U(1)_Y$ gauge theory with
two Higgs-doublets $\Phi_a$,
with the same hypercharge $1/2$,
and with real vacuum expectation values (vevs)
\begin{equation}
\langle \Phi_a \rangle
=
\left(
\begin{array}{c}
0\\
v_a/\sqrt{2}
\end{array}
\right)\, , \ \ \ (a=1,2),
\label{vev}
\end{equation}
with $v = \sqrt{v_1^2 + v_2^2} = 246\textrm{GeV}$.
Our definition for the charge is $Q=T_3 + Y$,
and we introduce the angle $\beta$ through
\be
\tan{\beta}= v_2/v_1\, .
\label{eq:beta}
\ee

The most general 2HDM scalar potential may be written as
\begin{eqnarray}
V_H
&&
=\, 
m_{11}^2 \Phi_1^\dagger \Phi_1 + m_{22}^2 \Phi_2^\dagger \Phi_2
- \left[ m_{12}^2 \Phi_1^\dagger \Phi_2 + \mathrm{H.c.} \right]
\nonumber\\[2pt]
&&
+\, \tfrac{1}{2} \lambda_1 (\Phi_1^\dagger\Phi_1)^2
+ \tfrac{1}{2} \lambda_2 (\Phi_2^\dagger\Phi_2)^2
+ \lambda_3 (\Phi_1^\dagger\Phi_1) (\Phi_2^\dagger\Phi_2)
\nonumber\\[2pt]
&&
+\, \lambda_4 (\Phi_1^\dagger\Phi_2) (\Phi_2^\dagger\Phi_1)
+ \left[
\tfrac{1}{2} \lambda_5 (\Phi_1^\dagger\Phi_2)^2
+ \mathrm{H.c.}
\right]
\nonumber\\[2pt]
&&
+ \left[
\lambda_6 (\Phi_1^\dagger\Phi_1) (\Phi_1^\dagger\Phi_2)
+ \lambda_7 (\Phi_2^\dagger\Phi_2) (\Phi_1^\dagger\Phi_2)
+ \mathrm{H.c.}
\right],
\label{VH1}
\end{eqnarray}
where ``H.c.''~stands for Hermitian conjugation.
The coefficients
$m_{11}^2$, $m_{22}^2$, and $\lambda_1,\cdots,\lambda_4$
are real while
$m_{12}^2$, $\lambda_5$, $\lambda_6$ and $\lambda_7$
may be complex.

When extending this model to the fermion sector, one finds
flavour changing neutral scalar interactions, which are very
strongly constrained by experiments on neutral meson systems.
This problem can be solved by imposing a $Z_2$ symmetry:
$\Phi_1 \rightarrow \Phi_1$; $\Phi_2 \rightarrow - \Phi_2$
\cite{Glashow:1976nt,Paschos:1976ay}.
If the symmetry is exact, then
the quadratic term $m_{12}^2 = 0$,
and the quartic terms $\lambda_6 = \lambda_7=0$.
(The quartic term $\lambda_5$ can then be made real by
a simple rephasing of $\Phi_2$.)
This has the consequence that the model has no decoupling limit.
That is, one cannot make arbitrarily large the masses of the new
particles resultant from the presence of the second scalar doublet,
thus approaching smoothly the SM limit.
Such a decoupling is a desirable feature,
especially since the couplings probed by current
LHC data are consistent with the SM predictions,
within errors of order 20\% \cite{Khachatryan:2016vau}.
Decoupling is recovered by reintroducing
$m_{12}^2 \neq 0$ \cite{Gunion:2002zf},
which,
because it breaks softly the $Z_2$ symmetry,
does not affect the renormalizability of the theory.

Most articles addressing this model then make the choice
that $m_{12}^2$  and $\lambda_5$ are both real and the vevs are real,
arguing that CP conservation in the scalar sector has been imposed
(choice 1).
Then one would proceed to discuss the various implementations of the
$Z_2$ symmetry in the fermion sector, and perform a variety of
fits to experiment.
Among these, one must fit the well measured CP violation with origin
in the Cabibbo-Kobayashi-Maskawa (CKM) matrix
\cite{Cabibbo:1963yz,Kobayashi:1973fv},
accommodated by the complex Yukawa couplings (choice 2).
The problem is that choice 1 and choice 2 are incompatible.

Indeed, either one applies the CP symmetry to the whole Lagrangian,
in which case the Yukawa couplings are real and one cannot account
for the observed CP violation; or, else, one does \textit{not}
apply CP symmetry anywhere, allowing the Yukawa couplings to be
complex, but then allowing also for a complex
$m_{12}^2$.\footnote{Recall that $\lambda_5$ can always
be made real through a convenient rephasing of $\Phi_2$.
In fact, one could instead use the rephasing to make $m_{12}^2$ real,
at the price of getting a complex $\lambda_5$.
What really matters is the rephasing invariant quantity
$\textrm{Im}\left[ \lambda_5^\ast \left(m_{12}^2\right)^2 \right]$.
For simplicity, we will make the discussion in the basis where $\lambda_5$
is real.}
Said otherwise,
requiring complex CKM and excluding the
parameter $\textrm{Im}\left( m_{12}^2\right)$
leads to a non-renormalizable theory.
At sufficiently high loop level, the CP violation
in the quark sector will leak into the scalar sector,
through a divergent contribution that cannot be absorbed by
a $\textrm{Im}\left( m_{12}^2\right)$ counterterm,
which was absent from the theory from the start.

So why do all articles fitting the real 2HDM ``model'' to experiment
ignore this problem?
Because the divergent contribution can only be shown to happen 
in at least
three loops.
However,
precisely because they are divergent, the problem cannot be ignored
if one wishes to use a theoretically sound model.

Given the fact that
the problem seems to occur due to (the lack of)
$\textrm{Im}\left(m_{12}^2\right)$,
one could be tempted to assume that such a dimension two operator
could not affect renormalizability.
And indeed, it cannot affect renormalizability due to its soft-breaking
of the $Z_2$ symmetry. But the problem with CP symmetry being
invoked is \textit{not}
that it is broken by $m_{12}^2$ (real or complex); rather,
it is (hard) broken
by the dimension four Yukawa couplings.

It is true that one can look at the real 2HDM as a limiting case of
the $Z_2$ 2HDM, softly broken by a complex $m_{12}^2$.
This model is known as the complex 2HDM (C2HDM) and has been
studied in detail; see, for example,
\cite{Ginzburg:2004vp,ElKaffas:2006gdt,Arhrib:2010ju,Barroso:2012wz,
Inoue:2014nva,Fontes:2014xva,Grzadkowski:2014ada,Fontes:2017zfn,Boto:2020wyf,Cheung:2020ugr}.
In that case, one can choose any tree-level values for the parameters,
and, in particular, set $\textrm{Im}\left( m_{12}^2 \right) = 0$
at tree level.
In that context, 
setting $\textrm{Im}\left( m_{12}^2 \right) = 0$ at tree level,
does not constitute a problem,
since the theory does have its counterterm and is renormalizable.
Is this the same as the real 2HDM? No:
setting $\textrm{Im}\left( m_{12}^2 \right) = 0$ in the C2HDM means that
we are studying a very specific corner of tree-level parameter space
of a more general model. The real 2HDM, where there is no
$\textrm{Im}\left( m_{12}^2 \right)$ nor its counterterm, is not a consistent model.

There is a more physical way to state the non renormalization problem.
In any 2HDM there are three neutral scalars ($h_1$, $h_2$, and $h_3$),
and a pair of charged scalars $H^\pm$.
Typically, it is assumed that the lightest neutral scalar ($h_1$)
coincides with the 125GeV particle found at LHC.\footnote{This is not mandatory.
One can accommodate the possibility that the 125GeV particle is not
the lightest neutral scalar, but we shall not concern ourselves here
with that case. See for example \cite{Ferreira:2012my,Bernon:2015wef}.}
In the real 2HDM,
the (proclaimed) lack of CP violation in the scalar sector,
leads to the separation of the three neutral scalars into
one single CP odd scalar ($A$)
and two CP even (usually denoted by $h$ for the lightest and $H$ for the heaviest).
If the CKM CP violation indeed seeps into the scalar sector,
then there should be divergent contributions to the $h$-$A$ and $H$-$A$
two-point functions.
There will also be divergent contributions to the A tadpole.
Since such terms are absent from the real scalar sector at tree level,
there are no counterterms to absorb those infinities,
and the theory is formally inconsistent.

\subsection{The scalar sector}

We start from the scalar sector of the real 2HDM
\begin{eqnarray}
V_r
&&
=\, 
m_{11}^2 \Phi_1^\dagger \Phi_1 + m_{22}^2 \Phi_2^\dagger \Phi_2
- m_{12}^2 \left[\Phi_1^\dagger \Phi_2 + \Phi_2^\dagger \Phi_1 \right]
\nonumber\\[2pt]
&&
+\, \tfrac{1}{2} \lambda_1 (\Phi_1^\dagger\Phi_1)^2
+ \tfrac{1}{2} \lambda_2 (\Phi_2^\dagger\Phi_2)^2
+ \lambda_3 (\Phi_1^\dagger\Phi_1) (\Phi_2^\dagger\Phi_2)
\nonumber\\[2pt]
&&
+\, \lambda_4 (\Phi_1^\dagger\Phi_2) (\Phi_2^\dagger\Phi_1)
+ 
\tfrac{1}{2} \lambda_5
\left[
(\Phi_1^\dagger\Phi_2)^2 + (\Phi_2^\dagger\Phi_1)^2
\right],
\label{Vreal}
\end{eqnarray}
with all parameters real, and we parametrize the fields in the original basis as
\begin{eqnarray}
\Phi_1 &=&
\left(
\begin{array}{c}
c_\beta G^+ - s_\beta H^+\\*[1mm]
\tfrac{1}{\sqrt{2}}
\left[
v c_\beta + (c_\alpha H - s_\alpha h) + i (c_\beta G^0 - s_\beta A)
\right]
\end{array}
\right),
\nonumber\\*[2mm]
\Phi_2 &=&
\left(
\begin{array}{c}
s_\beta G^+ + c_\beta H^+\\*[1mm]
\tfrac{1}{\sqrt{2}}
\left[
v s_\beta + (s_\alpha H + c_\alpha h) + i (s_\beta G^0 + c_\beta A)
\right]
\end{array}
\right).
\label{Phis_physical}
\end{eqnarray}
Throughout $c_\theta=\cos{\theta}$ and $s_\theta=\sin{\theta}$,
for whatever angle $\theta$ appears as a sub-index.
Comparing \eqref{vev} and \eqref{Phis_physical},
we recognize the choice $v_1 = v c_\beta$ and $v_2= v s_\beta$.
With that choice,
$G^+$ and $G^0$ will be massless and $H^+$ is the physical charged scalar
of mass $m_{H^\pm}^2$.\footnote{This is in fact a feature of the general
2HDM (which can be extended to multiple doublets) related to the existence
of a ``Higgs basis'' \cite{Botella:1994cs} -- $H_1=v_1^* \Phi_1 + v_2^* \Phi_2$,  $H_2=-v_2 \Phi_1 + v_1 \Phi_2$ -- where all the vev is in the first
doublet~\cite{Georgi:1978ri,Donoghue:1978cj}.}
These facts are confirmed by
substituting \eqref{Phis_physical} in \eqref{Vreal}.
Performing the expansion, one sees that there is no linear term in $A$
(and, thus, no possibility to absorb any infinities that might appear in
A tadpoles at loop level),
nor are there any quadratic $hA$ or $HA$ terms
(and, similarly, no possibility to absorb any infinities that might appear in
the corresponding two-point functions at loop level).

The expansion does contain linear (tadpole) terms 
for H ($t_H$) and h ($t_h$).
Equating these tadpoles to zero, one obtains the same conditions that one would
obtain by finding the stationarity conditions
$\partial V_r/\partial v_1=0$ and $\partial V_r/\partial v_2=0$
directly from \eqref{Vreal}.
Those two equations can be solved for $m_{11}^2$ and $m_{22}^2$,
which are then substituted back into the expression for the potential.
After this substitution using the vacuum conditions,
there are no quadratic $G^0G^0$ and $G^+ G^-$ terms.
There are also no mixed $G^0 A$ or $G^\pm H^\mp$ terms,
meaning that as expected $G^0$ and $G^+$ are Goldstone bosons,
while $A$ and $H^+$ are already mass eigenstates.
One finds quadratic terms of the type $HH$, $hh$, and $hH$.
The angle $\alpha$ is chosen to kill the latter,
meaning that $h$ and $H$ are the physical fields.
Using $v=246\textrm{GeV}$ and $m_h=125\textrm{GeV}$,
the scalar sector of the real 2HDM is parametrized by six further parameters:
the mixing angles $\alpha$ and $\beta$;
the masses $m_H$, $m_A$, $m_{H^\pm}$;
and the soft-breaking parameter $m_{12}^2= \textrm{Re}\left( m_{12}^2\right)$.

\subsection{CP violation from the CKM matrix}

In the SM and in the real 2HDM, CP violation arises from the complex
Yukawa couplings.
When the quark fields are rotated into their mass basis,
all CP violation phases are contained in the 
CKM matrix \cite{Cabibbo:1963yz,Kobayashi:1973fv}.
But, each quark field can still be rephased at will,
thus moving the CP violating phase around the various entries
of the CKM matrix.
The only rephasing invariant quantity is
\cite{Jarlskog:1985ht,Jarlskog:1985cw,Dunietz:1985uy}
\be
I^{\alpha i}_{\beta  j}
= \textrm{Im}\left(
V_{\alpha i} V_{\beta j} V_{\alpha j}^\ast V_{\beta i}^\ast
\right)\, ,
\label{eq:Jarlskog_1}
\ee
where $\alpha \neq \beta$ and $i \neq j$.
We use the notation of \cite{Branco:1999fs},
where Greek letters $\alpha, \beta, \gamma, \dots$ refer to up-type quarks
$u_\alpha = u, c, t$,
while Roman letters $i, j, k, \dots$
refer to down-type quarks
$d_i = d, s, b$.
There are nine distinct four quark combinations with different flavours:
$(ds)$, $(db)$, and $(sb)$ for the down-type quarks
times the three for up-type quarks: $(uc)$, $(ut)$, and $(ct)$.
Using the unitarity of the CKM matrix,
the following symmetries hold
\be
I^{\alpha i}_{\beta  j} = I^{\beta  j}_{\alpha i}
=
- I^{\alpha j}_{\beta  i} = - I^{\beta i}_{\alpha  j},
\ee
showing that indeed there is only one independent CP violating quantity.
The antisymmetry with respect to interchange of same quark-type indices
is easiest to see in the form 
\be
I^{\alpha i}_{\beta  j}
= J\, \sum_{\gamma, k} \epsilon_{\alpha \beta \gamma}\, \epsilon_{i j k}\, ,
\label{eq:Jarlskog_2}
\ee
where
$J$ is the Jarlskog invariant
\cite{Jarlskog:1985ht,Jarlskog:1985cw,Dunietz:1985uy},
defined for example as 
$J=I^{ud}_{cs}$.
Notice that
$I^{ud}_{cs} \neq 0$
 even though only quarks from the first two families
are involved.
This does not contradict the fact that there would be no CP violating
phase in the SM if there existed only two families of quarks.
The fact 
that
 the CKM matrix is $3 \times 3$ unitary
(and, thus, has one irremovable complex phase) is built into
eq.~(\ref{eq:Jarlskog_2}).
Nevertheless, in CP violating processes with no external quarks
(and, thus, a quark loop), the appearance of four CKM ($V$) factors
can only occur in diagrams at the three loop level and above.

As emphasized by Khriplovich and Pospelov \cite{Pospelov:1991zt}
and by Booth \cite{Booth:1993af}
in the context of the electric dipole moments (edm)
of the $W$ and the electron, the antisymmetry
of $I^{\alpha i}_{\beta j}$ is very powerful.
Any CP violating amplitude from a fermion box diagram will appear as the product
$I^{\alpha i}_{\beta  j}$ with some amplitude
\be
\mathcal{A}(m_{u_\alpha},m_{u_\beta},m_{d_i},m_{d_j})\, .
\label{eq:A_ms}
\ee
When all contributions are summed over ($\alpha$, $\beta$, $i$, and $j$),
all terms in $\mathcal{A}(m_{u_\alpha},m_{u_\beta},m_{d_i},m_{d_j})$
symmetric under $\alpha \leftrightarrow \beta$,
or $i \leftrightarrow j$ will not contribute.
A much more involved analysis along these lines
was used in \cite{Pospelov:1991zt,Booth:1993af}
to show that the SM electroweak contributions to the
electric dipole moments
of the $W$  and the electron vanish to two-loop
and three-loop approximation, respectively.

\section{A theoretically unsound toy model}
\label{sec:toymodel}

To better illustrate our claim, we consider here a
toy model that suffers from the same inconsistency as the real 2HDM.
In both models, CP conservation is enforced in a particular
sector of an otherwise CP violating theory.
As a consequence, CP violating radiative effects end up contaminating
the alleged CP conserving sector, thus leading to divergences
that cannot be absorbed by the counterterms.
The major feature of our toy model is that such divergences show up
immediately at one-loop order.
Therefore, it constitutes a simple materialization of the
same theoretical pathology that we claim to be present
at three-loop or above in the real 2HDM.

The present model is inspired in a model by Pilaftsis~\cite{Pilaftsis:1998pe},
which, however, does not suffer from the flaw we wish to point out.
Consider two Abelian gauge symmetries $U(1)_Q$ and $U(1)_B$,
with gauge bosons $A_{\mu}$ and $B_{\mu}$,
respectively.
Suppose also four complex scalars, $\Phi_1$,
$\Phi_2$, $\chi_L$ and $\chi_R$, with charges 
\be
Q(\Phi_1) = 0, \,
Q(\Phi_2) = 0, \,
Q(\chi_L) = 1, \,
Q(\chi_R) = 1,
\ee
\be
B(\Phi_1) = 1, \,
B(\Phi_2) = 1, \,
B(\chi_L) = -\dfrac{1}{5}, \,
B(\chi_R) = \dfrac{4}{5},
\ee
where $Q$ and $B$ represent the conserved charges of $U(1)_Q$ and $U(1)_B$,
respectively.
A discrete symmetry $D$ is imposed on the fields, under which:
\be
\Phi_1 \stackrel{D}{\to} - \Phi_1, \ \ 
\Phi_2 \stackrel{D}{\to} \Phi_2, \ \
\chi_L \stackrel{D}{\to} - \chi_L, \ \
\chi_R \stackrel{D}{\to} \chi_R.
\ee
However, $D$ is allowed to be softly broken.
The complete renormalizable Lagrangian can then be written in four terms,
\be
\mathcal{L} = \mathcal{L}_{kin} + \mathcal{L}_{\Phi} + \mathcal{L}_{\chi} + \mathcal{L}_{\Phi\chi},
\label{eq:main}
\ee
where $\mathcal{L}_{kin} $ represents the kinetic
terms\footnote{We assume no $A_\mu$-$B_\mu$ kinetic mixing.} and
\bs
\bea
- \mathcal{L}_{\Phi} &=& \mu_1^2 \Phi_1^* \Phi_1 +
\mu_2^2 \Phi_2^* \Phi_2 + \mu^2 \Phi_1^* \Phi_2 +
(\mu^2)^* \, \Phi_2^* \Phi_1 \nonumber\\
&& + \lambda_1 {\left ( \Phi_1^{*} \Phi_1 \right )}^2
+ \lambda_2 {\left ( \Phi_2^{*} \Phi_2 \right )}^2
+ \lambda_{34} \, \Phi_1^{*} \Phi_1 \Phi_2^{*} \Phi_2  \nonumber\\
&& + \lambda_5 { \left (  \Phi_1^{*} \Phi_2 \right )}^2
+ \lambda_5^* {\left ( \Phi_2^{*} \Phi_1 \right )}^2,
\label{eq:pot} \\[2mm]
- \mathcal{L}_{\chi} &=&  m_L^2 \, \chi_L \chi_L^*
+ m_R^2 \, \chi_R \chi_R^* + \rho_1 (\chi_L^* \chi_L)^2 \nonumber\\
&& + \rho_2 (\chi_R^* \chi_R)^2
+  \rho_{34} \,  \chi_L^* \chi_L  \chi_R^* \chi_R,\label{eq:7} \\[2mm]
- \mathcal{L}_{\Phi\chi} &=&  f_1 \, \Phi_1  \chi_L \chi_R^*
+ f_1^* \, \Phi_1^*  \chi_L^* \chi_R
+ f_2 \, \Phi_2  \chi_L \chi_R^* \nonumber\\
&& + f_2^* \, \Phi_2^*  \chi_L^* \chi_R
+ g_1 \, \Phi_1^{*} \Phi_1 \chi_L^* \chi_L
+ g_2 \, \Phi_2^{*} \Phi_2\chi_L^* \chi_L \nonumber \\
&&  + \, g_3 \, \Phi_1^{*} \Phi_1 \chi_R^* \chi_R
+ \, g_4 \, \Phi_2^{*} \Phi_2 \chi_R^* \chi_R.
\label{eq:Yuk}
\eea
\es
The parameters $\mu^2$, $\lambda_5$, $f_1$ and $f_2$ are
in general complex, while the remaining ones are real by
construction. The terms involving $\mu^2$ and $f_2$
break the symmetry $D$ softly.
It is easy to show that the conditions for CP conservation are:
\bs
\label{eq:conds}
\bea
\text{Im} \left[ \mu^2 \, f_1 \, f_2^* \right] &=& 0, \label{eq:9a}\\
\text{Im}  \Big[ \lambda_5 \, f_1^2  \,(f_2^*)^2 \Big] &=& 0, \label{eq:9b}\\
\text{Im}  \Big[ \lambda_5^* \, (\mu^2)^2 \Big] &=& 0.
\label{eq:9c}
\eea
\es
Mimicking the usual real 2HDM treatment,
we take $\langle \Phi_1 \rangle$ and $\langle \Phi_2 \rangle$ real
and parametrize
\bs
\label{eq:6}
\bea
\Phi_1 &=& 
\tfrac{1}{\sqrt{2}} \left( v_1 + H_1 + i A_1 \right),\\
\Phi_2 &=& 
\tfrac{1}{\sqrt{2}} \left( v_2 + H_2 + i A_2 \right),
\eea
\es
where 
$v_1$, $v_2$ are real and non-negative,
and
$H_1$, $H_2$, $A_1$, and $A_2$ are real fields.
The vevs $v_1$ and $v_2$ break spontaneously the gauge symmetry $U(1)_B$.
Recall that $\mu^2$, $\lambda_5$, $f_1$, $f_2$ are in general complex.

But suppose we force CP to be conserved in $\mathcal{L}_{\Phi}$
by the ad-hoc imposition that
$\mu^2$ and $\lambda_5$ are real.
This we will do in the following.
It will lead
to irremovable divergences at one-loop, as we
now show.

We start by determining the minimization (or tadpole) equations.
These are:\footnote{The tadpole equations for $A_1$ and $A_2$ are trivially zero.}
\bs
\label{eq:tree-tads}
\bea
0 &=& t_{H_1} := \dfrac{\partial \mathcal{L}_{\Phi}}{\partial
H_1}\bigg|_{< >=0}
\no 
&=& -v_1 \left( \mu_1^2 + \dfrac{v_2}{v_1} \mu ^2
+  v_1^2 \lambda_1 + \dfrac{1}{2} v_2^2 \lambda _{34}
+  v_2^2 \, \lambda _5\right)\hs {-1mm}, \\
0 &=& t_{H_2}  := \dfrac{\partial \mathcal{L}_{\Phi}}{\partial
H_2}\bigg|_{< >=0}
\no
&=& -v_2 \left( \mu_2^2 + \dfrac{v_1}{v_2} \mu ^2
+  v_2^2 \lambda_2 + \dfrac{1}{2} v_1^2 \lambda _{34}
+  \, v_1^2 \, \lambda _5\right) \hs {-1mm},
\eea
\es
where  $t_{H_1}, t_{H_2}$ represent the tree-level
tadpoles for $H_1$, $H_2$, respectively, and
$< >=0$ means that the expectation
values of all fields on the right hand side of
eqs.~(\ref{eq:6}) are set to zero.
Recall that we are taking $\mu^2$ and $\lambda_5$ as
real parameters. For that reason, the mass matrices for $H_1$ and
$H_2$, on the one hand, and $A_1$ and $A_2$, on the other,
can be separately diagonalized. We thus define the
angles $\theta$ and $\beta$ such that:
\bs
\bea
\left(\begin{array}{c}
H_{1} \\
H_{2}
\end{array}\right)
&=&
\left(\begin{array}{cc}
c_{\theta} & -s_{\theta} \\
s_{\theta} & c_{\theta}
\end{array}\right)\left(\begin{array}{l}
h \\
H
\end{array}\right),
\\
\left(\begin{array}{c}
A_{1} \\
A_{2}
\end{array}\right)
&=&
\left(\begin{array}{cc}
c_{\beta} & -s_{\beta} \\
s_{\beta} & c_{\beta}
\end{array}\right)\left(\begin{array}{c}
G^0 \\
A
\end{array}\right),
\eea
\es
where $h$ and $H$ are the CP even states, and
$A$ and $G^0$ are the CP odd states, where
$G^0$ is the massless would-be Goldstone boson.\footnote{$G^0$
is eaten by the longitudinal component of $B_{\mu}$ in the
unitary gauge through the Higgs mechanism.
The fact that $G^0$ is massless forces $\beta$ to
obey the relation $\tan \beta = \frac{v_2}{v_1}$.}
We continue using the short notation $s_x \equiv \sin(x)$,
$c_x \equiv \cos(x)$, for a generic angle $x$.
As for the $\mathcal{L}_{\chi}$ sector, the mass matrix is given by
\be
-\mathcal{L}^{\chi}_{\text{mass}}=
\left(\begin{array}{cc}
\chi_L^* & \chi_R^*
\end{array}\right)
\left(\begin{array}{cc}
a & b^* \\
b & c
\end{array}\right)
\left(\begin{array}{c}
\chi_L \\
\chi_R
\end{array}\right),
\label{eq:mymass}
\ee
with
\bs
\bea
a &=& \frac{1}{2} g_1 v_1^2+\frac{1}{2} g_2 v_2^2+m_L^2, \\
b &=& \dfrac{f_1 v_1+ f_2 v_2}{\sqrt{2}}, \\
c &=& \frac{1}{2} g_3 v_1^2+\frac{1}{2} g_4 v_2^2+m_R^2\, ,
\eea
\es
where $a$ and $c$ are real, while $b$ is in principle complex.
However, we can rephase $\chi_R$ through
$\chi_R \to e^{i \, \text{arg}(b)} \chi_R$
so that it absorbs the phase of $b$.
In the new basis, then, $b$ is real, which implies that 
the mass matrix in eq.~(\ref{eq:mymass}) is
symmetric.\footnote{If it were hermitian, one would
need a unitary matrix to diagonalize it, instead of
an orthogonal one. Moreover, note that 
$b$ real forces the relation
$f_1^* = f_1 + \left( f_2 - f_2^*\right)  \tan{\beta}$.}
We thus need an orthogonal matrix with a new angle
$\phi$ to diagonalize the states:
\be
\left(\begin{array}{c}
\chi_L \\
\chi_R
\end{array}\right)
=
\left(\begin{array}{cc}
c_{\phi} & -s_{\phi} \\
s_{\phi} & c_{\phi}
\end{array}\right)\left(\begin{array}{c}
\chi_1 \\
\chi_2
\end{array}\right),
\ee
where $\chi_1$ and $\chi_2$ are the (complex)
diagonalized states with (real) masses $M_1$ and $M_2$. 

\begin{figure}[htp] 
\hspace{6mm}
\centering 
\begin{subfigure}{2.5cm} 
\begin{fmffile}{200} 
\begin{fmfgraph*}(70,40) 
\fmfset{arrow_len}{3mm} 
\fmfset{arrow_ang}{20} 
\fmfleft{nJ1} 
\fmflabel{$A$}{nJ1} 
\fmfright{nJ2} 
\fmf{dashes,tension=2}{nJ1,nJ1J1J4} 
\fmf{phantom,tension=3}{nJ2,nJ2J3J2} 
\fmf{scalar,label=$\chi_1$,right=1}{nJ2J3J2,nJ1J1J4} 
\fmf{scalar,label=$\chi_1$,right=1}{nJ1J1J4,nJ2J3J2} 
\end{fmfgraph*} 
\end{fmffile} 
\end{subfigure} 
\hspace{9mm} 
\begin{subfigure}{2.5cm} 
\begin{fmffile}{201} 
\begin{fmfgraph*}(70,40) 
\fmfset{arrow_len}{3mm} 
\fmfset{arrow_ang}{20} 
\fmfleft{nJ1} 
\fmflabel{$A$}{nJ1} 
\fmfright{nJ2} 
\fmf{dashes,tension=2}{nJ1,nJ1J1J4} 
\fmf{phantom,tension=3}{nJ2,nJ2J2J3} 
\fmf{scalar,label=$\chi_2$,right=1}{nJ2J2J3,nJ1J1J4} 
\fmf{scalar,label=$\chi_2$,right=1}{nJ1J1J4,nJ2J2J3} 
\end{fmfgraph*} 
\end{fmffile} 
\end{subfigure}
\caption{Feynman diagrams contributing to
the one-loop tadpole of the CP odd state.}
\label{fig:mytads}
\end{figure}
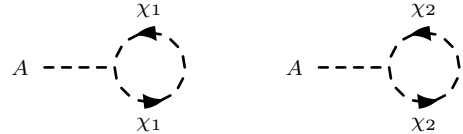

When considering the theory up to one-loop level,
one should renormalize it in order to obtain finite
$S$-matrix elements. This is done through the usual procedure:
taking an independent set of parameters, identifying them
with bare quantities (represented in the following with
the index $0$) and relating them to their renormalized
equivalents through a counterterm.
Tadpoles can be taken care of through the tadpole scheme identified by PRTS in \cite{Denner:2019vbn}.
It then follows from the set of Eqs.~(\ref{eq:tree-tads}) that:
\bs
\label{eq:loop-tads}
\bea
\delta t_{H_1} &:=& -v_1 \bigg( \mu_{1,0}^2 + \dfrac{v_2}{v_1} \mu_0 ^2 +  v_1^2 \lambda_{1,0} + \dfrac{1}{2} v_2^2 \lambda _{34,0} \no
&& \hs{20mm} + v_2^2 \, \lambda_{5,0}\bigg),  \\
\delta t_{H_2} &:=& -v_2 \bigg( \mu_{2,0}^2 + \dfrac{v_1}{v_2} \mu_0 ^2 +  v_2^2 \lambda_{2,0} + \dfrac{1}{2} v_1^2 \lambda_{34,0} \no
&& \hs{20mm} + v_1^2 \, \lambda_{5,0}\bigg),
\eea
\es
where the tadpole counterterms $\delta t_{H_1}$
and $\delta t_{H_2}$ are such that
$\delta t_{H_1}  = -T_{H_1}$ and
$\delta t_{H_2}  = -T_{H_2}$, with $T_{H_1}$ and  $T_{H_2}$
being the one-loop tadpole for $H_1$ and $H_2$, respectively.
The set of \eqref{eq:loop-tads} fixes
the values for $v_1$ and $v_2$ at one-loop level.
Note that, since we imposed CP conservation in
$\mathcal{L}_{\Phi}$, there are no tadpole counterterms
for the CP odd fields.
Specifically, in the mass basis,
\be
\delta t_{A} = 0.
\label{eq:tads-mass-odd}
\ee
But it is easy to see that this is inconsistent.
Indeed, there is a one-loop tadpole for $A$,
whose diagrams are represented in Fig.~\ref{fig:mytads}.
The sum of diagrams is divergent.
In fact,
\be
\left(T_{A}\right)\Big|_{\text{div}} =
- \dfrac{1}{\varepsilon}
\dfrac{{c_{\phi }}  \, {s_{\phi }} \,
\left( M_1^2 - M_2^2 \right)  \,
\text{Im}[f_2]}{8 \, \sqrt{2} \, \pi^2 \, c_{\beta}},
\label{eq:taddivs}
\ee
in $d=4-2\varepsilon$ dimensions,
where $T_{A}$ represents the one-loop tadpole for
$A$ and $\big|_{\text{div}}$ means that we consider
only divergent parts.

The origin of the problem lies in the fact that we
imposed $\mu^2$ and $\lambda_5$ to be real.
To clarify this point, let us provisionally
take these parameters to be complex, as they originally were.
By rewriting eq.~(\ref{eq:pot}) in terms of bare quantities,
and separating the real and imaginary parts of
$\mu_0^2$ and $\lambda_{5,0}$,
the terms proportional to these parameters are:
\be
\begin{split}
&- \mathcal{L}_{\Phi_0}
\ni
\mu_0^2 \, \Phi_{1,0}^* \Phi_{2,0}
+ (\mu_0^2)^* \, \Phi_{2,0}^* \Phi_{1,0} \\
& \hs{10mm} +
\lambda_{5,0} \, \left(\Phi_{1,0}^* \Phi_{2,0}\right)^2
+ \lambda_{5,0}^* \, \left(\Phi_{2,0}^* \Phi_{1,0}\right)^2
\\
&= \text{Re} [\mu_0^2]
(\Phi_{1,0}^* \Phi_{2,0} + \Phi_{2,0}^* \Phi_{1,0}) \\
& \hspace{5mm} +
\text{Re} [\lambda_{5,0}] \Big\{\left(\Phi_{1,0}^* \Phi_{2,0}\right)^2
+ \left(\Phi_{2,0}^* \Phi_{1,0}\right)^2\Big\} \\
& \hspace{5mm} +
i \, \text{Im} [\mu_0^2] \left(\Phi_{1,0}^* \Phi_{2,0}
- \Phi_{2,0}^* \Phi_{1,0}\right)\\
& \hspace{5mm} + i \,
\text{Im} [\lambda_{5,0}] \Big\{\left(\Phi_{1,0}^*
\Phi_{2,0}\right)^2 - \left(\Phi_{2,0}^* \Phi_{1,0}\right)^2\Big\}.
\end{split}
\label{eq:decide}
\ee
As a consequence, when we set
$\text{Im} [\mu_0^2] = \text{Im} [\lambda_{5,0}] = 0$,
we are not including in the model the terms of the two
last lines of eq.~(\ref{eq:decide}).
Naturally, since such terms are not in the model,
there is no counterterm for the parameters involved therein.
That is, there is neither $\text{Im} [\delta \mu^2]$
nor $\text{Im} [\delta \lambda_5]$.\footnote{The situation
would not be different if we decided
to exclude any other term from the theory. For example,
had we decided not to include the term
proportional to $\lambda_1$ in the model,
there would be no counterterm $\delta \lambda_1$.}
Now, it is a matter of course that this would not be a problem
if the fact that we did not include the terms in the last two
lines of eq.~(\ref{eq:decide}) would follow from a symmetry that forbade them.
In other words, should there be a symmetry in the theory that
proscribed those terms, they could logically not be included;
and since the symmetry would prevent any Green's functions
generated by such terms from showing up, there would never be
divergences involved therein, so that the absence of
counterterms for them would never be a problem.
So, for example, if CP was a symmetry of the theory,
it would preclude those terms, in which case the absence
of $\text{Im} [\delta \mu^2]$ and $\text{Im} [\delta \lambda_5]$
would not be inconsistent.

However, CP is \textit{not} a symmetry of theory:
even if we try to impose it in the $\mathcal{L}_{\Phi}$ sector,
it still is violated in the $\mathcal{L}_{\Phi \chi}$
sector through the phases of $f_1$ and $f_2$,
as Eqs.~(\ref{eq:9a}) and (\ref{eq:9b}) show.
So, there is no CP symmetry forbidding the terms in
the last two lines of eq.~(\ref{eq:decide}).
As a consequence, even if we exclude them, CP violating radiative effects
can nonetheless contribute to the Green's functions involved therein.
Such Green's functions will in general be divergent;
but since the last two lines of eq.~(\ref{eq:decide})
were not included in the theory, there will in general
not be enough counterterms to absorb them.

\begin{figure}[htp] 
\centering 
\hspace{6mm}
\begin{subfigure}{2.5cm} 
\begin{fmffile}{3} 
\begin{fmfgraph*}(60,30) 
\fmfset{arrow_len}{3mm} 
\fmfset{arrow_ang}{20} 
\fmfleft{nJ1} 
\fmflabel{$A,G_0$}{nJ1} 
\fmfright{nJ2} 
\fmflabel{$h,H$}{nJ2} 
\fmf{dashes,tension=3}{nJ1,nJ1J2J3} 
\fmf{dashes,tension=3}{nJ2,nJ2J1J4} 
\fmf{scalar,label=$\chi_1$,right=1}{nJ1J2J3,nJ2J1J4} 
\fmf{scalar,label=$\chi_2$,right=1}{nJ2J1J4,nJ1J2J3} 
\end{fmfgraph*} 
\end{fmffile} 
\end{subfigure} 
\hspace{17mm} 
\begin{subfigure}{2.5cm} 
\begin{fmffile}{4} 
\begin{fmfgraph*}(60,39) 
\fmfset{arrow_len}{3mm} 
\fmfset{arrow_ang}{20} 
\fmfleft{nJ1} 
\fmflabel{$A,G_0$}{nJ1} 
\fmfright{nJ2} 
\fmflabel{$h,H$}{nJ2} 
\fmf{dashes,tension=3}{nJ1,nJ1J1J4} 
\fmf{dashes,tension=3}{nJ2,nJ2J3J2} 
\fmf{scalar,label=$\chi_2$,right=1}{nJ2J3J2,nJ1J1J4} 
\fmf{scalar,label=$\chi_2$,right=1}{nJ1J1J4,nJ2J3J2} 
\end{fmfgraph*} 
\end{fmffile} 
\end{subfigure}
\\[8mm]
\hspace{6mm}
\begin{subfigure}{2.5cm} 
\begin{fmffile}{1} 
\begin{fmfgraph*}(60,30) 
\fmfset{arrow_len}{3mm} 
\fmfset{arrow_ang}{20} 
\fmfleft{nJ1} 
\fmflabel{$A$}{nJ1} 
\fmfright{nJ2} 
\fmflabel{$h,H$}{nJ2} 
\fmf{dashes,tension=3}{nJ1,nJ1J1J4} 
\fmf{dashes,tension=3}{nJ2,nJ2J3J2} 
\fmf{scalar,label=$\chi_1$,right=1}{nJ2J3J2,nJ1J1J4} 
\fmf{scalar,label=$\chi_1$,right=1}{nJ1J1J4,nJ2J3J2} 
\end{fmfgraph*} 
\end{fmffile} 
\end{subfigure} 
\hspace{17mm} 
\begin{subfigure}{2.5cm} 
\begin{fmffile}{2} 
\begin{fmfgraph*}(60,30) 
\fmfset{arrow_len}{3mm} 
\fmfset{arrow_ang}{20} 
\fmfleft{nJ1} 
\fmflabel{$A$}{nJ1} 
\fmfright{nJ2} 
\fmflabel{$h,H$}{nJ2} 
\fmf{dashes,tension=3}{nJ1,nJ1J1J4} 
\fmf{dashes,tension=3}{nJ2,nJ2J2J3} 
\fmf{scalar,label=$\chi_1$,right=1}{nJ2J2J3,nJ1J1J4} 
\fmf{scalar,label=$\chi_2$,right=1}{nJ1J1J4,nJ2J2J3} 
\end{fmfgraph*} 
\end{fmffile} 
\end{subfigure} 
\caption{Feynman diagrams contributing to the
one-loop CP violating 2-point functions.}
\label{fig:myf}
\end{figure}
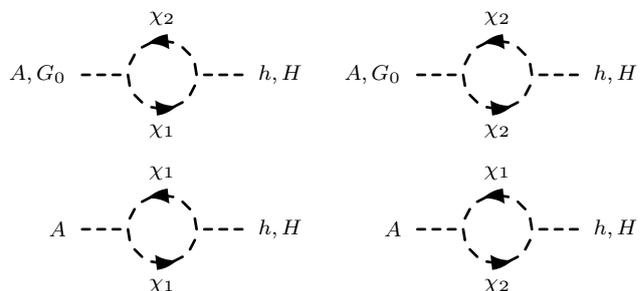

We have already seen one example of Green's function
that indeed cannot be renormalized: the one-loop 1-point
function $T_{A}$. Other examples are the one-loop
CP violating 2-point functions $\Sigma^{G^0 h}$,
$\Sigma^{G^0 H}$, $\Sigma^{A h}$ and $\Sigma^{A H}$
for the scalar-pseudoscalar mixing of $G^0 \, h$,
$G^0 \, H$, $A\, h$ and $A\, H$, respectively.
Their Feynman diagrams are represented in Fig.~\ref{fig:myf}. 
Just like in the case of $T_{A}$, there simply is no
counterterm for these functions, which nonetheless
are divergent. Their divergent parts are:
\bs
\label{eq:divs}
\bea
&& \Sigma^{G^0 h}(k^2)\Big|_{\text{div}}
= \Sigma^{AH}(k^2)\Big|_{\text{div}}
\nonumber \\[1.5mm]
&& \hs{7mm} = - \dfrac{1}{\varepsilon}
\, \sin(\beta  - \theta) \,
\frac{\text{Im}\left[f_2\right]
\left( f_1 + f_2 \tan \beta \right)}{16 \pi ^2},
\\[3mm]
&& \Sigma^{Ah}(k^2)\Big|_{\text{div}}
= - \Sigma^{G^0 H}(k^2)\Big|_{\text{div}}
\nonumber \\[1.5mm]
&& \hs{7mm} =
- \dfrac{1}{\varepsilon} \, \cos(\beta  - \theta)
\, \frac{\text{Im}\left[f_2\right]
\left( f_1 + f_2 \tan \beta \right)}{16 \pi ^2}.
\eea
\es
In conclusion, the fact that we imposed
$\mu^2$ and $\lambda_5$ 
to be real leads to several
divergences that cannot be removed by counterterms.

There are two ways to heal this model: either CP is
imposed as a whole, or it is not imposed at all.
In the first case, all the three relations in
eqs.~(\ref{eq:conds}) should be verified, which implies
that there is a basis where $\mu^2$, $\lambda_5$,
$f_1$ and $f_2$ are all real. In this scenario, therefore,
CP violating Green's functions are precluded, which implies,
in particular, that no divergent CP violating Green's
functions will appear in any order.
This is consistent with what we obtained in
eqs.~(\ref{eq:taddivs}) and (\ref{eq:divs}),
which vanish in the limit of real $f_1$ and $f_2$.
In the second case, $\mu^2$, $\lambda_5$, $f_1$ and $f_2$
are in general complex parameters, which implies
that their counterterms are also in general complex.
Since CP is violated, there are no scalar states with
well-defined CP, and Green's functions will in general
be CP violating.
The model is renormalizable as long as all the terms
compatible with the symmetries are included.
Finally, note that, in such a CP violating scenario,
there may be regions of the parameter space in which 
$\lambda_5$ and $\mu^2$ are real, and $f_1$ and $f_2$ complex.
But this is a completely different situation from that where one
builds a theory taking \textit{ab initio} $\lambda_5$ and
$\mu^2$ real, while $f_1$ and $f_2$ in general complex.
In fact, while the former situation corresponds to a
particular solution of a consistent, renormalizable theory,
the latter suffers from the inconsistencies we have shown.

\section{\label{sec:3looptad}Three-loop tadpole for $A$ in the real 2HDM}

Our goal is to check whether the complex phases of the fermion
mixing matrices introduce CP violating effects into the
otherwise CP conserving scalar sector of the real 2HDM via radiative corrections.
For this purpose, we focus on the effects of quark-mixing.
The quantity that signifies quark-induced CP violation in a
convention independent way
is the Jarlskog invariant $J$ in eq.~(\ref{eq:Jarlskog_2}).
So, we are looking for radiative corrections to the 2HDM
which contain this quantity.
As the simplest check,
we have looked for diagrams proportional to $J$,
contributing to the $A$ tadpole.
As argued above,
this can only happen in amplitudes with at least four vertices,
each containing a factor of $V_{u_\alpha d_j}$ and additionally,
a vertex to couple to $A$ (none of the $A$-couplings have
CKM-factors).
Therefore, the first possible appearance of $J$ in
$A$-tadpoles is at three loops.
This is indeed what we find at the amplitude-level.

\begin{figure}[htp] 
	\hspace{6mm}
	\centering 
	\begin{subfigure}{2.5cm} 
		\begin{fmffile}{three-loop-tadpole-J1}
			\fmfset{thin}{.7pt}
			\fmfset{dash_len}{1.5mm}
			\fmfset{arrow_len}{2.5mm}
			\fmfset{wiggly_len}{2.2mm}
			\fmfset{arrow_len}{2.5mm} 
			\begin{fmfgraph*}(70,50) 
				\fmfleft{l} 
				\fmflabel{\footnotesize{$A$}}{l} 
				\fmfright{r}
				\fmftop{t}
				\fmfbottom{b}
				\fmf{phantom}{vrb,r,vrt}
				\fmf{phantom}{vlb,l,vlt}
				\fmf{phantom,tension=1.5}{vlt,t,vrt}
				\fmf{phantom,tension=1.5}{vlb,b,vrb}
				\fmf{phantom}{r,vr,vl,l}
				\fmf{dashes,tension=2}{l,vl}
				\fmffreeze
				\fmf{fermion,label.side=left,label.dist=3,label=$u_\alpha$}{vlb,vl,vlt}
				\fmf{fermion}{vlt,vrt}
				\fmf{fermion,left=0.5}{vrt,vrb}
				\fmf{fermion}{vrb,vlb}
				\fmf{dashes_arrow,right=0.5,label.side=left,label.dist=2,label=\tiny{$H^-$}}{vlb,vlt}
				\fmf{wiggly,right=0.5,label.side=left,label.dist=2,label=\tiny{$W^{\vphantom{-}}$}}{vrt,vrb}
				\fmfv{label=\footnotesize{$d_j$},label.angle=-90,label.dist=0.}{t}
				\fmfv{label=\footnotesize{$d_i$},label.angle=90,label.dist=0.}{b}
				\fmfv{label=\footnotesize{$u_\beta$},label.angle=180,label.dist=0.05}{r}
			\end{fmfgraph*} 
		\end{fmffile}
	\caption{}
	\end{subfigure} 
	\hspace{9mm} 
	\begin{subfigure}{2.5cm} 
		\begin{fmffile}{three-loop-tadpole-J2}
			\fmfset{thin}{.7pt}
			\fmfset{dash_len}{1.5mm}
			\fmfset{arrow_len}{2.5mm}
			\fmfset{wiggly_len}{2.2mm}
			\fmfset{arrow_len}{2.5mm} 
			\begin{fmfgraph*}(70,50) 
				\fmfleft{l} 
				\fmflabel{\footnotesize{$A$}}{l} 
				\fmfright{r}
				\fmftop{t}
				\fmfbottom{b}
				\fmf{phantom}{vrt,r,vrb}
				\fmf{phantom}{vlt,l,vlb}
				\fmf{phantom,tension=1.5}{vlb,t,vrb}
				\fmf{phantom,tension=1.5}{vlt,b,vrt}
				\fmf{phantom}{r,vr,vl,l}
				\fmf{dashes,tension=2}{l,vl}
				\fmffreeze
				\fmf{fermion,label.side=right,label.dist=3,label=$u_\alpha$}{vlb,vl,vlt}
				\fmf{fermion}{vlt,vrt}
				\fmf{fermion,right=0.5}{vrt,vrb}
				\fmf{fermion}{vrb,vlb}
				\fmf{dashes_arrow,left=0.5,label.side=right,label.dist=2,label=\tiny{$H^-$}}{vlb,vlt}
				\fmf{wiggly,left=0.5,label.side=right,label.dist=2,label=\tiny{$W^{\vphantom{-}}$}}{vrt,vrb}
				\fmfv{label=\footnotesize{$d_j$},label.angle=-90,label.dist=0.}{t}
				\fmfv{label=\footnotesize{$d_i$},label.angle=90,label.dist=0.}{b}
				\fmfv{label=\footnotesize{$u_\beta$},label.angle=180,label.dist=0.05}{r}
			\end{fmfgraph*} 
		\end{fmffile} 
	\caption{}
	\end{subfigure}
	\caption{Example of pair of Feynman diagrams where $J$ factorizes (for fixed $\alpha,\beta,i,j$). They differ only by the direction of fermion flow (or equivalently by the exchange of the two
	down-type quarks, $d_i \leftrightarrow d_j$).}
	\label{fig:jarl-pair}
\end{figure}
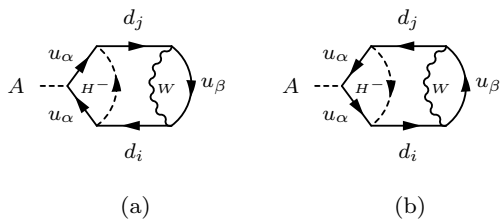

An example of a pair of diagrams yielding the
Jarlskog invariant is shown in Fig.~\ref{fig:jarl-pair}.
If contributions of this kind were divergent, then one would lack the
respective counterterm needed to absorb these divergences within the real 2HDM.
Therefore, we checked for the existence of a leading $1/\varepsilon^3$-pole
in said contributions, which would be a strong indication for the
necessity of a genuine three-loop $A$-tadpole counterterm.
For this purpose, we generated all three-loop $A$-tadpole amplitudes
$T_A$ with the only condition being that a fixed set
of quarks $\{u_\alpha, u_\beta, d_i, d_j\}$ must be contained.
Other contributions are CP conserving operators and, thus,
irrelevant to our discussion.

Our calculation was carried out in three independent ways
(two numeric; one analytical),
fully explained in section~\ref{sec:details}.
The result is\footnote{Notice that the angle $\beta$
in $s_\beta c_\beta$ is the angle in eq.~(\ref{eq:beta}),
while in all other instances of eq.~(\ref{eq:tadpole}),
$\beta$ refers to the up-type quark being considered.
Here and henceforth, which $\beta$ is meant should be
clear from the context.
}
\begin{fmffile}{tadpole-A0}
	\fmfset{thin}{.7pt}
	\fmfset{dash_len}{1.5mm}
	\begin{align}
	\left(T_A\right)^{\alpha i}_{\beta j} &= 
	-i \,\Big( \;
	\begin{gathered}
	\vspace{-4pt}
	\begin{fmfgraph*}(30,40)
	\fmfright{t}
	\fmfleft{b}
	\fmf{dashes,label=\footnotesize{$A$},label.dist=2}{b,t}
	\fmfblob{14}{t}
	\end{fmfgraph*}
	\end{gathered}
	\quad
	\Big)^{u_\alpha d_i}_{u_\beta d_j}
	\nonumber \\
	&=
	\frac{g^5}{8 \varepsilon^3 m_W^3 s_{\beta} c_{\beta}}
	M^{\alpha i}_{\beta j}\, I^{\alpha i}_{\beta j}
	+\mathcal{O}(\varepsilon^{-2}),
	\label{eq:tadpole}
	\end{align}
\end{fmffile}%
where there is no sum over repeated indices, and
\begin{align}
M^{\alpha i}_{\beta j}
=&
(m_{u_\alpha}^2 - m_{u_\beta}^2) (m_{d_i}^2 - m_{d_j}^2) \nonumber \\
&\times(m_{u_\alpha}^2 - m_{d_i}^2 + m_{u_\beta}^2 - m_{d_j}^2)\, .
\label{eq:M}	
\end{align}
The fact that such different calculational techniques
yielded the same result is truly significant.

Remarkably,
when summing over all different sets
of up- and down-type quark contributions,
the leading pole vanishes exactly.
Indeed,
it is easy to show that summing the combination
$M^{\alpha i}_{\beta j} I^{\alpha i}_{\beta j}$
over all the nine distinct sets of four different
quarks (two up-type and two down-type),
the result vanishes.
Notice that
both $M^{\alpha i}_{\beta j}$ and $I^{\alpha i}_{\beta j}$
are antisymmetric under $\alpha \leftrightarrow \beta$
(or $i \leftrightarrow j$).
Thus, the vanishing of 
eq.~(\ref{eq:tadpole}) is not due to the simple symmetry reasons
mentioned in connection with eq.~(\ref{eq:A_ms}).
It is the specific form of the mass term
$M^{\alpha i}_{\beta j}$ in eq.~(\ref{eq:M}) which makes this possible.
We cannot see how one would have guessed from the start
this rather peculiar mass combination.
We resonate with Khriplovich and Pospelov's remark in
the context of edm that:
``We cannot get rid of the feeling that this simple
result (...) should have a simple transparent explanation.
Unfortunately, we have not been able to find it.''

But the physical consequence is quite clear:
\begin{equation} \label{eq:sum-over-poles}
\sum_{\alpha < \beta} \sum_{i<j}
\left(T_{A}\right)^{\alpha i}_{\beta j}
=
\mathcal{O}(\varepsilon^{-2}).
\end{equation}
It remains uncertain whether this cancellation has a physical origin
or it is to be interpreted as \emph{accidental}.
There is the possibility that the next order
$1/\varepsilon^2$-poles would be non-vanishing.
Otherwise, a genuine CP violating
tadpole counterterm for $A$ would only become relevant
at the four-loop level.

\section{Details of the calculation}
\label{sec:details}

In this section, we discuss our derivation of \eqref{eq:tadpole}.
First note that a complete calculation of the renormalized three-loop tadpole for $A$ would require
the full renormalization of the model at both one-loop and two-loop order.
Although unlikely, one cannot exclude the possibility that combinations of the one- and two-loop counterterms of a CP-conserving scalar sector conspire to cancel the divergences of a CP-violating three-loop tadpole.
Secondly, one caveat in our calculation is the treatment of amplitudes
with an uneven number of $\gamma$-matrices together
with $\gamma_5$.
We chose to work in \emph{naive dimensional regularisation}
with the expectation that the leading $\varepsilon$-poles do
not depend on the choice of a $\gamma_5$-scheme.
This claim is supported by the findings in~\cite{Belusca-Maito:2020ala} at the one-loop level.

As mentioned before, at least three generations of quarks are necessary to generate a CP-violating tadpole.
Therefore, we focus on a particular set of diagrams.
Let us then define $S^{\{dcbt\}}$ as the set of all the three-loop tadpole diagrams for $A$ containing the quarks $d$, $c$, $b$, $t$.
We started by generating the amplitudes for $S^{\{dcbt\}}$ in an $R_\xi$-gauge.
We did this through two independent softwares: \texttt{FeynMaster} \cite{Fontes:2019wqh} (which makes use of \texttt{FeynRules}~\cite{Christensen:2008py} and 
\texttt{QGRAF}~\cite{Nogueira:1991ex}) and \texttt{FeynArts} \cite{Hahn:2000kx}.\footnote{It is worth emphasizing that the three-loop tadpole amplitudes generated with \texttt{FeynArts} and \texttt{FeynMaster} coincide.}
At three loops, there are 360 amplitudes containing the quarks $d,c,b$ and $t$.
However, 120 among them involve two closed loops of fermions, which means that they can never factorize the Jarlskog invariant $J$; and since a) the tadpole for $A$ violates $CP$ and b) all the $CP$-violation in the real 2HDM must be proportional to $J$ (with $J$ being the only rephasing-invariant quantity signifying CP-violation), those diagrams must sum up to zero. 
We checked this explicitly using \texttt{FeynCalc}~\cite{Mertig:1990an,Shtabovenko:2016sxi,Shtabovenko:2020gxv}.

We then focused on the remaining 240 diagrams.
After simplifying the Lorentz and Dirac algebra of the 240 diagrams with \texttt{FeynCalc}, another 32 diagrams, such as the ones with two internal W-boson loops, vanish immediately (in naive dimensional regularisation) due to the chirality of the interactions involved.
This eventually left us with 208 diagrams which can be categorized as follows:
\begin{enumerate}
	\item The first group of diagrams can be generated from connecting $A$ to any fermion line in Fig.~\ref{fig:category-1} and the corresponding ones with reversed fermion flow giving 16 diagrams.
	The same goes for all possible vector boson and scalar insertions, namely $$\{HW,WH,HG,GH,HH,GW,WG,GG\}.$$
	Diagrams where we connect $A$ to a line with an attached $W$-loop vanish though, leaving us with $8\times 16 - 4 \times 4 = 112$ diagrams.
\begin{figure}[htp] 
	\hspace{6mm}
	\centering 
	\begin{subfigure}{2.5cm} 
		\begin{fmffile}{three-loop-tadpole-C1a}
			\fmfset{thin}{.7pt}
			\fmfset{dash_len}{1.5mm}
			\fmfset{arrow_len}{2.5mm}
			\fmfset{wiggly_len}{2.2mm}
			\fmfset{arrow_len}{2.5mm} 
			\fmfframe(0,10)(0,10){
				\begin{fmfgraph*}(60,50) 
					\fmfleft{lb,lt} 
					\fmfright{rb,rt}
					\fmf{fermion,label.side=left,label=$c$}{lb,lt}
					\fmf{fermion,label.side=left,label=$d$}{lt,rt}
					\fmf{fermion,label.side=left,label=$t$}{rt,rb}
					\fmf{fermion,label.side=left,label=$b$}{rb,lb}
					\fmf{dashes_arrow,right=0.6,label.side=right,label.dist=2,label=\tiny{$H^-$}}{lb,lt}
					\fmf{wiggly,right=0.6,label.side=left,label.dist=2,label=\tiny{$W^{\vphantom{-}}$}}{rt,rb}
				\end{fmfgraph*}
			}
		\end{fmffile} 
		\caption{}
	\end{subfigure} 
	\hspace{9mm} 
	\begin{subfigure}{2.5cm} 
		\begin{fmffile}{three-loop-tadpole-C1b}
			\fmfset{thin}{.7pt}
			\fmfset{dash_len}{1.5mm}
			\fmfset{arrow_len}{2.5mm}
			\fmfset{wiggly_len}{2.2mm}
			\fmfset{arrow_len}{2.5mm} 
			\fmfframe(0,10)(0,10){ 
				\begin{fmfgraph*}(60,50) 
					\fmfleft{lb,lt} 
					\fmfright{rb,rt}
					\fmf{fermion,label.side=left,label=$d$}{lb,lt}
					\fmf{fermion,label.side=left,label=$t$}{lt,rt}
					\fmf{fermion,label.side=left,label=$b$}{rt,rb}
					\fmf{fermion,label.side=left,label=$c$}{rb,lb}
					\fmf{dashes_arrow,right=0.6,label.side=right,label.dist=2,label=\tiny{$H^+$}}{lb,lt}
					\fmf{wiggly,right=0.6,label.side=left,label.dist=2,label=\tiny{$W^{\vphantom{-}}$}}{rt,rb}
				\end{fmfgraph*} 
			}
		\end{fmffile}
		\caption{}
	\end{subfigure}
	\caption{Attaching $A$ to the fermion lines of these diagrams and the ones with reversed fermion flow generates the first category of relevant diagrams.}
	\label{fig:category-1}
\end{figure}
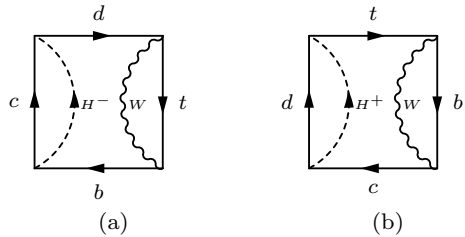

	\item An example of the second group is shown in Fig.~\ref{fig:category-2}. From this diagram and the one with reversed fermion flow, we get 8 diagrams by cyclic permutations of the fermions.
	The $A$ can be connected to either $\{WH,HW,GH,HG\}$ and we can have either a $W$-, $H$-, or $G$-loop in the diagram. 
	This gives $3 \times 4 \times 8 = 96$ diagrams.
\begin{figure}[htp] 
	\hspace{6mm}
	\centering 
	\begin{fmffile}{three-loop-tadpole-C2a}
		\fmfset{thin}{.7pt}
		\fmfset{dash_len}{1.5mm}
		\fmfset{arrow_len}{2.5mm}
		\fmfset{wiggly_len}{2.2mm}
		\fmfset{arrow_len}{2.5mm} 
		\fmfframe(0,10)(0,10){
			\begin{fmfgraph*}(80,60) 
				\fmfleft{l} 
				\fmflabel{\footnotesize{$A$}}{l} 
				\fmfright{r}
				\fmftop{t}
				\fmfbottom{b}
				\fmf{phantom}{vrb,r,vrt}
				\fmf{phantom}{vlb,l,vlt}
				\fmf{phantom,tension=1.5}{vlt,t,vrt}
				\fmf{phantom,tension=1.5}{vlb,b,vrb}
				\fmf{phantom}{r,vr,vl,l}
				\fmf{dashes,tension=2}{l,vl}
				\fmffreeze
				\fmf{wiggly,label.side=left,label=$W$}{vl,vlt}
				\fmf{dashes_arrow,label.side=left,label=$H^-$}{vlb,vl}
				\fmf{fermion,label.side=left,label=$d$}{vlt,vrt}
				\fmf{fermion,label.side=right,label=$t$}{vrt,vrb}
				\fmf{fermion,label.side=left,label=$b$}{vrb,vlb}
				\fmf{fermion,label.side=right,label=$c$}{vlb,vlt}
				\fmf{wiggly,left=0.5,label.side=left,label=\footnotesize{$W$}}{vrt,vrb}
			\end{fmfgraph*} 
		}
	\end{fmffile} 
	\caption{The second set of relevant diagrams is generated from permutations of the fermions in this diagram and by replacing the $W$-insertions with a Goldstone boson.}
	\label{fig:category-2}
\end{figure}
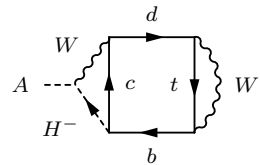
\end{enumerate}

We proceeded to numerically evaluate the most divergent part of the 208 diagrams using \texttt{FIESTA}~\cite{Smirnov:2015mct} in Feynman gauge, \ie at $\xi_W=1$.
In order to generate input integrals, the FeynCalc function \texttt{ApartFF} was essential for decomposing the diagrams via partial fractioning.
This decomposition yielded scalar integrals for which we could easily get an accurate result from \texttt{FIESTA}.
One integral type needed additional attention though, namely the one with a scalar product in the numerator and five different propagator factors (see~\ref{sec:problem-integral} for a discussion).
This type of integral yielded large error estimates in \texttt{FIESTA}, such that the results could no be trusted.
In order to obtain an exact result for those, we used integration-by-parts identities with \texttt{FIRE}~\cite{Smirnov:2019qkx} to decompose this integral type into a set of scalar integrals.
The intermediate steps required to link \texttt{FeynCalc}, \texttt{FIRE} and \texttt{FIESTA} were performed by two independent sets of private codes.

For the numerical input values of the scalar sector, one should choose a point in parameter space which does not violate any theoretical or experimental constraints.
The theoretical bounds include boundedness from below, perturbative
unitarity~\cite{Kanemura:1993hm,Akeroyd:2000wc,Ginzburg:2005dt} as well
as electroweak precision measurements using the oblique parameters S,
T and U~\cite{Branco:2011iw}.  The experimental constraints include the
exclusion bounds from Higgs searches at LHC that were verified using
\texttt{HiggsBounds}~\cite{Bechtle:2013wla,Bechtle:2020pkv} and the signal
strengths for the SM-like Higgs boson were forced to be within
2$\sigma$ of the fits given in~\cite{Khachatryan:2016vau,Aad:2019mbh}.
Among the points that pass all constraints, we pick the following one:
\be
\begin{split}
	\alpha &= -0.83797 \\
	\beta &= 0.73908 \\
	m_{H^{\pm}} &= 581.18 \, \, \text{GeV}, \\
	m_H &= 592.81 \, \, \text{GeV}, \\
	m_{A} &= 597.44 \, \, \text{GeV}, \\
	m_{12}^2 &= 19.458 \, \, \text{TeV}.
\end{split}
\label{eq:mypoint}
\ee
Using these together with \eqref{eq:input-SM}, the result for the most divergent part of $S^{\{dcbt\}}$ is:
\be
\left(i T_{A}\right)^{cd}_{tb} 
= 2392.6 \, \, (\text{GeV})^3
 \times \dfrac{1}{\varepsilon^3}  J + \mathcal{O}(\varepsilon^{-2}).
\label{eq:TAdiv1}
\ee
\n We tried to ascertain whether sets of diagrams with different combinations of quarks could possibly cancel with each other by checking the result for another set of quarks.
For example, consider the set of diagrams $S^{\{dcst\}}$, defined as identical to $S^{\{dcbt\}}$ except that the $b$ quark is replaced by an $s$ quark.
Using the same point in parameter space (eq.~\ref{eq:mypoint}), the result for the most divergent part of $S^{\{dcst\}}$ is
\be
\left(i T_{A}\right)^{cd}_{ts}
 = - 0.91341 \, \, (\text{GeV})^3
\times \dfrac{1}{\varepsilon^3} \, J + \mathcal{O}(\varepsilon^{-2}).
\label{eq:TAdiv2}
\ee
Clearly, the numbers differ, which might lead one to believe that summing over the results for all quarks combinations would likely yield a non-zero result.

Only later, our attention was drawn to \texttt{TVID}~\cite{Bauberger:2017nct,Bauberger:2019heh}, a software package for the numerical evaluation of arbitrary three-loop vacuum integrals -- see also \cite{Martin:2016bgz}.
The authors discuss a set of three master integrals into which any three-loop tadpole diagram can be decomposed.
Using the same amplitude decomposition as for our numerical evaluation, we were then able to map to this set of master-integrals and use the analytic pole expressions of \texttt{TVID} (or equivalently of Ref.~\cite{Freitas:2016zmy}) to acquire the result for the $1/\varepsilon^3$-pole shown in \eqref{eq:tadpole}.\footnote{This calculation also showed that the evaluation of scalar three-loop integrals with up to five propagators and different mass scales via \texttt{FIESTA} yields accurate results for the leading poles. This might be the first such stress test on this package.}
After confirming that the analytic results coincided with our previous numerical findings, we were able to sum over all possible quark combinations using \eqref{eq:tadpole} to find the surprising result of the poles cancelling, as shown in \eqref{eq:sum-over-poles}.

\section{\label{sec:conclusions}Conclusions}

We argued that the real 2HDM 
	can suffer from theoretical inconsistencies,
as the simultaneous enforcement of CP conservation in the potential and allowance of CP violation in another sector 
	may lead
to divergences that cannot be removed by the counterterms.
Because such divergences cannot show up at two-loop level and below, the unsoundness of the model has been by and large ignored in the literature.
But the problem cannot be dismissed.
In order to highlight its
	potential theoretical unsoundness
we introduced a simple toy model, characterized by the same inconsistency as the real 2HDM.
There, and as we showed, the irremovable divergences (that are expected at least at three-loops in the real 2HDM) show up immediately at one-loop level.
This simple example ought to make the point: the real 2HDM 
	could
suffer from the same kind of pathology.
We addressed this claim by calculating the leading pole of the three-loop one-point function of the $A$ field in the real 2HDM.
We showed that, surprisingly, the pole vanishes exactly after summing all contributions.
This does not mean that the model is sound after all, but only that its unsoundness is 
	likely
to be found either at lower order in $1/\varepsilon$ or upon two-loop renormalization, or possibly at four-loop order.
	A complete discussion would require the full one- and two-loop renormalization as well as a discussion of the $\gamma_5$-scheme beyond naive dimensional regularisation though.

We hope that our work spurs further interest in this subject and that a full calculation will become possible in the future.

\begin{acknowledgements}
We are grateful to Francisco Botella, Renato Fonseca,
Paulo Nogueira and Howard Haber for discussions,
as well as to Vladyslav Shtabovenko and Alexander Smirnov for useful
suggestions concerning softwares for multi-loop evaluation.

\n This work is supported in part
by the Portuguese \textit{Funda\c{c}\~{a}o para a Ci\^{e}ncia e Tecnologia}
(FCT) under contracts UIDB/ 00777/2020, UIDP/00777/2020,
CERN/FIS-PAR/0004/2017, and PTDC/FIS-PAR/29436/2017.
D.F. is also supported by the Portuguese \textit{Funda\c{c}\~ao para a Ci\^encia e Tecnologia} under the project SFRH/BD/135698/2018.
M.L. is supported partially by the DFG Collaborative Research Center
TRR 257 “Particle Physics Phenomenology after the
Higgs Discovery”.

\end{acknowledgements}

\appendix

\section{Integral decomposition}\label{sec:problem-integral}
As mentioned in section~\ref{sec:details}, one integral type in our amplitude decomposition needed special attention.
The one in question is
\begin{align}
	&U_5^{(1,2)} (m_1,m_2,m_3,m_4,m_5) \nonumber \\ 
	&= i \frac{e^{3\gamma_E \varepsilon}}{\pi^{3d/2}}\int \! \! \mathrm{d}^d q_1 \, \mathrm{d}^d q_2 \, \mathrm{d}^d q_3 \frac{q_1 \cdot q_2}{(q_1^2 - m_1^2) (q_2^2 - m_2^2) (q_3^2 - m_3^2)} \nonumber \\
	& \qquad \times\frac{1}{  \big((q_1-q_3)^2-m_4^2\big) \big((q_2- q_3)^2 - m_5^2\big) }.
\end{align}
Evaluating integrals of this kind with \texttt{FIESTA} yields small yet non-zero error estimates already for the leading pole which is why the numerical results could not be trusted.
Therefore, we made use of the integration by parts routines of \texttt{FIRE} to decompose $U_5^{(1,2)}$ into a set of scalar integrals, which in turn yield vanishing error estimates for the leading poles when evaluated with \texttt{FIESTA}.
Using analytic expressions for the leading poles of the resulting integral decomposition also made it possible to recombine everything into a joint expression for the leading pole of $U_5^{(1,2)}$, \viz
\begin{align}
		 U_5^{(1,2)}& (m_1,m_2,m_3,m_4,m_5) \nonumber \\
		=& \frac{1}{24 \varepsilon^3}
		\Big[
		(m_1^2 + m_2^2)^2 + (m_2^2 + m_5^2)^2 + (m_3^2 + m_4^2)^2 \nonumber \\
		& \quad \quad + (m_3^2 + m_5^2)^2 -4(m_2^2-m_3^2)^2 +m_2^4 + m_3^4
		\Big] \nonumber \\
		&  + \mathcal{O}(\varepsilon^{-2}).
\end{align}

\section{Numerical input values}

\be
\begin{split}
	m_W  &=  80.358 \, \, \text{GeV}, \\
	m_u  &=  2.2 \times 10^{-3} \, \, \text{GeV}, \\
	m_d  &=  4.8 \times 10^{-3} \, \, \text{GeV}, \\
	m_c  &=  1.4464 \, \, \text{GeV}, \\
	m_s  &=  0.093 \, \, \text{GeV}, \\
	m_t  &=  172.5 \, \, \text{GeV}, \\
	m_b  &=  4.8564 \, \, \text{GeV}, \\
	e  &=  0.30812, \\
	\sin \theta_w &= 0.47206.
\end{split}
\label{eq:input-SM}
\ee


\end{document}